\documentclass[conference]{IEEEtran}
\IEEEoverridecommandlockouts

\usepackage{graphicx,times,amsmath,amssymb,multirow,caption,subcaption,color}
\usepackage[LGR,T1]{fontenc}
\usepackage[latin9]{inputenc}
\usepackage{array,textcomp,stackrel,url,mathtools,enumerate}
\usepackage{multirow} %For multirow in tables
\usepackage{multicol} %For multicol in tables
\usepackage{booktabs,helvet,courier,float,csquotes}
\usepackage{algorithm,algcompatible}
\usepackage{cite}
\usepackage{flushend} % Makes the columns on the last page equal in length.
\def\BibTeX{{\rm B\kern-.05em{\sc i\kern-.025em b}\kern-.08em
    T\kern-.1667em\lower.7ex\hbox{E}\kern-.125emX}}

\usepackage{tikz}  \usetikzlibrary{matrix,chains,positioning,decorations.pathreplacing,arrows}
\algnewcommand\INPUT{\item[\textbf{Input:}]}%
\algnewcommand\OUTPUT{\item[\textbf{Output:}]}%

\makeatletter

\begin{document}
\title{Mixed Initiative Systems for Human-Swarm Interaction: Opportunities and Challenges}

% \author{Aya Hussein  and \and Hussein Abbass
%  \thanks{Authors are with the School of Engineering \& IT, University of New South Wales,
%  Canberra-Australia, (e-mail: a.hussein@student.adfa.edu.au ,h.abbass@adfa.edu.au).}% <-this % stops a space
% }
\author{
  Aya Hussein \\
    \IEEEauthorblockA{\textit{School of Engineering and Information Technology} \\
\textit{UNSW Canberra, Australia}
\\a.hussein@student.adfa.edu.au}
  
  \and Hussein Abbass\\
  \IEEEauthorblockA{\textit{School of Engineering and Information Technology} \\
\textit{UNSW Canberra, Australia}
\\h.abbass@adfa.edu.au}
}

\IEEEtitleabstractindextext{
}

\maketitle
\begin{abstract}

Human-swarm interaction (HSI) involves a number of human factors impacting human behaviour throughout the interaction. As the technologies used within HSI advance, it is more tempting to increase the level of swarm autonomy within the interaction to reduce the workload on humans. Yet, the prospective negative effects of high levels of autonomy on human situational awareness can hinder this process. Flexible autonomy aims at trading-off these effects by changing the level of autonomy within the interaction when required; with mixed-initiatives combining human preferences and automation's recommendations to select an appropriate level of autonomy at a certain point of time.  However,  the effective implementation of mixed-initiative systems raises fundamental questions on how to combine human preferences and automation recommendations, how to realise the selected level of autonomy, and what the future impacts on the cognitive states of a human are.
We explore open challenges that hamper the process of developing effective flexible autonomy. We then highlight the  potential benefits of using system modelling techniques in HSI by illustrating how they provide HSI designers with an opportunity to evaluate different strategies for assessing the state of the mission and for adapting the level of autonomy within the interaction to maximise mission success metrics. 
\end{abstract}

%\IEEEraisesectionheading{\section{Introduction}\label{intro}}
\section{Introduction}
%\IEEEPARstart{H}{uman} { 
Human Swarm Interaction (HSI) is a growing research field~\cite{wearable},~\cite{hsi} that deals with the interface between a human and a swarm of robots within a mission. This interface is not limited to the graphical user interface, but to the overall interaction interface including protocols for interactions. A swarm of robots consists of a group of individual robots, typically with limited processing capabilities~\cite{distributedConsensus}, whose local interactions can result in a complex behaviour~\cite{sharedControl}, e.g. flocking. The ability to generate complex behaviour from local interactions results in two advantageous properties: scalability and flexibility~\cite{scalableHSI}. In addition, the distributed nature of the swarm makes them robust to failures. Thus, swarm robotics have the potential to be used in different applications including material transformation~\cite{material}, agriculture~\cite{Agricultural}, urban search and rescue~\cite{usar}, monitoring and surveillance~\cite{visualFeedback}, and space exploration~\cite{space}. 

Nevertheless, the human element is necessary in swarm operations so as to make sure that swarm behaviours align with the objectives of the mission~\cite{sharedControl}. Moreover, humans are more capable of working in complex and dynamic environments. Thus, the interest in HSI has been growing in recent years~\cite{wearable}.

As humans become part of the interaction, it becomes crucial to consider human factors that affect human performance throughout the mission. Therefore, designing an interaction paradigm that takes human factors into consideration improves not only human performance but also the overall mission performance~\cite{FromHere}.
A realisation of such an interaction scheme exists in flexible autonomy systems in which task distribution and interface customisation can be contingent on the state of the mission including the human. Although systems with flexible autonomy have been shown to be superior to rigid systems with fixed autonomy, many aspects of flexible autonomy are still poorly understood. The limited understanding of these aspects can be partially attributed to the difficulty of setting human experiments for evaluating potential design options. 
In this work, we discuss how modelling systems with flexible autonomy can be beneficial for investigating the impacts of different automisation strategies on human states and mission performance.

The rest of the paper is organised as follows: we start by describing possible roles a human can take within a mission, in Section 2.  Then, in Section 3, we discuss human issues that can impact and be impacted by the state of the mission. Next, in Section 4, we discuss two properties of the swarm that are relevant to flexible autonomy. Next, we investigate how flexible autonomy can cater for effective interaction by changing the level of autonomy within a mission, in Section 5. In Section 6, we elaborate on a few open research problems confronting the formation of a concrete understanding of aspects of flexible autonomy as well as the potential benefits of using system modelling within the context of flexible autonomy in HSI. Finally, conclusions are given in Section 7. %} 

\section{Human Roles in HSI}
Many studies have shown that the human element can be beneficial or even essential for the success of swarm operations~\cite{hsi}~\cite{visualFeedback}. %~\cite{independentCamera}.  
While humans have superior cognitive abilities that enable them to deal with dynamic and unstructured environments, robots can perform specific and repetitive tasks precisely and quickly. Combining the capabilities of humans and robots can improve the success rate of complex missions.

However, humans can be assigned different roles in each mission. According to Scholtz~\cite{roles1}, there are five roles that humans can take in human-robot interactions (HRI). These roles are: supervisor, operator, teammate, bystander, and mechanic. As a supervisor, the human is mainly responsible for evaluating the overall situation against mission objectives. Thus a supervisor would be expected to get involved in mission level decisions or modify high level plans. On the other extreme, a human operator is required to control and monitor low level tasks on action level, such as the case of motor control. In such a case, robots are considered as extensions of humans' physical capabilities that can be remotely deployed in harsh or risky environments. For human teammates, they work with robots towards achieving mission goals. During the interaction, they can provide the robots with high level commands without modifying the overall goal. Obviously, there is no strict boundary between different human roles. For example, a human can work with the robot as a teammate, but may switch to a more supervisory role to modify a mission level plan or objective. 

Finally, and less related to our scope, come the roles of bystander and mechanic. A bystander doesn't interact explicitly with the robot, but their existence may result in changing low level actions, for example to avoid colliding with him/her. A mechanic interacts physically with the robot to modify abnormal behaviors or adjust its physical components.  

Previous research in HSI showed that humans perform better when they act as supervisors than  operators~\cite{meta}. For example, Kolling et al.~\cite{towards} found that in simulated foraging missions, even na\"ive
 robot swarms that act  in a fully autonomous mode outperform human operators controlling swarms low level actions. Nevertheless, they found that humans are better at adapting to unstructured environments. Given that real environments are usually complex and dynamic, the merits of involving a human in the loop to guide swarm operations can be seen. Thus, assigning more supervisory roles to the human could be a prudent choice to improve mission success.

\section{Human Factors in HSI}
%The previous discussion established an initial understanding of how roles can be distributed between the human and the swarm within a mission while leaving the details of these roles unspecified. It has been argued that humans should take on supervisory roles and the swarm should be assigned an appropriate level of autonomy based on its capacity. 
Discussing different roles of the human in HSI provides primitive guidelines on how humans can be designated within a mission, but leaves the details of task assignment unspecified. While the details of task distribution and interface design are mission specific, human factors within the interaction should be considered when designing a certain interaction scheme. The human factors community has devoted considerable efforts to studying human situational awareness, human workload, and human level of trust towards the automation as these factors and has identified that they are significant factors to human performance within a mission. The details of these factors are discussed in the following subsections.

\subsection{Situational Awareness}
Human interaction with a robot swarm can be proximate in which the human shares the same physical environment as the swarm, e.g~\cite{distributedConsensus}, or remote in which the human exists in a different environment and interacts with the swarm typically through a computer terminal~\cite{visualFeedback}. In both situations, the human has to maintain contextual and situational awareness and understanding of the current state of the swarm, the relevant aspects of the surrounding environment, and the progress of the mission. 

According to Endsley~\cite{sagat}, situational awareness (SA) is \enquote{the perception of the elements in the environment within a volume of time and space, the comprehension of their meaning, and the projection of their status in the near future}. Many studies have shown that SA has a significant impact on human decision making and ultimately on task performance~\cite{sagat}~\cite{FromHere}. For instance, Riley et al.~\cite{SAPerformance} found a significant positive correlation between human monitoring performance and their level of SA. Poor SA was found to be the reason behind many problems in robot assisted tasks~\cite{endsley1997}\cite{usar}~\cite{SAChapter}  %\cite{teaming}  
as it limits the human ability to detect and intervene to solve emergent problems. These studies raised the profile of SA in semi-autonomous systems. 

A widely used model for SA is Endsley's three-level model~\cite{sagat} in which the first level describes human knowledge of the state of relevant elements in the environment, the second level reflects the degree to which he/she integrates this data to understand the overall current situation, and the third level describes his/her ability to make relevant predictions in the near future.

A number of techniques have been used to measure human SA. Situation Awareness Global Assessment Technique (SAGAT)~\cite{sagat} is a widely used  knowledge-based method in which the human takes part in a simulated mission. The simulation gets frozen at random points of time so that the operator can answer some questions measuring different levels of SA. The provided answers can then be evaluated against the correct answers. Despite their ability to measure SA directly, knowledge-based techniques cannot be applied in real missions because they interrupt mission execution and because the correct answers are not known in advance.
 
A useful computational model for SA should consider the main factors that form and affect it. A taxonomy of factors affecting SA in HRI is provided  in~\cite{SAChapter}. These factors can belong to the task (e.g. its level of autonomy), the system (e.g. communication characteristics), the environment (e.g. complexity),  personal skills (e.g. experience and cognitive abilities), or the  interface ( e.g. level of information fusion). Furthermore, the dynamic nature of SA should also be captured, for example the relationship between SA and the level of workload on the human~\cite{endsley1997}.

\subsection{Workload}
Another important factor that influences human performance within a mission is the level of workload imposed on the human.
Firstly, the human ability to develop and maintain the desired level of SA was found to be affected by the level of workload. Endsley et al.~\cite{endsley1997} argued that in scenarios characterised by very high levels of workload that exceed human cognitive resources, humans may not be able to attend to all the available information, which can result in significant drop in the level of SA. 

Moreover, the effect of workload on human performance was investigated in many studies~\cite{WLReliance}~\cite{HATeaming}~\cite{visualFeedback}~\cite{2018}. Findings suggest that both very low levels and very high levels of workload can cause human performance degradation. Very low levels of workload can result in arousal decrements that causes "out-of-the-loop" problem~\cite{FromHere}. When the humans are "out-of-the-loop" they become more like observers than supervisors such that their ability to intervene to correct for system failures decrease substantially~\cite{Workload2017}. On the other hand, as the workload exceeds human cognitive capacity, human performance is expected to decline~\cite{abbass2014computational}. Thus, it is important that the workload is maintained within the acceptable range to increase the effectiveness of the human in the operation. 

Workload is a well understood human factor that has been receiving considerable attention from researchers in different fields. Early studies considered human workload as consisting of objective and subjective components~\cite{Campbell}. While the objective component consists of factors stemming from task structure, the interface, or the environment; the subjective component is made up of factors belonging to the human performer of the task including their experience, cognitive abilities, skills, and self confidence.

Studies on the objective factors of workload in HRI showed that workload is affected by the level of autonomy of the robot~\cite{usarJ}~\cite{LOA}, such that at low levels, the human becomes responsible for planning and executing low level actions which results in considerable workload~\cite{usarJ}. As the level of autonomy increases, human functions are not eliminated, but the nature of human tasks becomes more supervisory. For these supervisory roles, the human becomes responsible for monitoring the performance and making mission-level decisions that may require him/her to attend to and integrate large amounts of information~\cite{swarmingNetworks}. 

In addition, the number of interruptions (e.g, alerts or threats) and the frequency of task switching can increase the workload imposed on the human. Interruptions can hinder the smooth execution of the task at hand~\cite{Workload2017} and increase the probability of errors~\cite{Feigh}. It has been shown that it can take humans a long time to recover from interruptions and restore main-task related SA~\cite{HATeaming}. Consistent task switching has also been associated with significantly slower responses and higher errors~\cite{HATeaming}. Thus, multitasking can incur substantial increases in workload~\cite{Workload2017}, particularly when the similarity between tasks increases in terms of their presentation or demands on similar cognitive resources~\cite{MRT}.

A number of computational models has been proposed for describing workload in HRI and related fields. In~\cite{DESWorkload}, Donmez et al. proposed a model for a human supervising multiple heterogeneous unmanned vehicles (UVs) using discrete event simulation (DES). In their model, the human is represented as a serial server that processes a queue of tasks which can be generated from the UVs or from the external environment. A model for server characteristics was used in which the human attention allocation strategy and the effects of the level of workload on attention efficiency are used to determine server performance in terms of service time. The level of autonomy of the UVs is represented as the rate of arrival of UV tasks. Rusnock et al.~\cite{Rusnock1} also used DES to model cognitive workload in HRI in military applications. They found a high correlation between the predicted levels of workload and those measured by task load index (NASA-TLX).

The subjective component of workload has received much less interest in HRI and HSI. However, there are some evidences that some skills and traits can mitigate the subjective workload on the human. These skills include: effective use of working memory~\cite{endsley1997}, attention allocation~\cite{SAChapter} and multi-tasking~\cite{ucav},  task related experience~\cite{SAChapter}, and spatial abilities~\cite{Jentsch}.
A variety of psycho-physiological measures like electroencephalography (EEG) and heart-rate variability  has been used to assess the overall workload on the human. A recent review on different techniques for workload assessment can be found in~\cite{workloadSurvey}.    

\subsection{Trust}
In their seminal paper~\cite{trustDef}, Lee et al. defined  human trust in an agent as \enquote{the attitude that an agent will help achieve an individual's goals in a situation characterized by uncertainty and vulnerability}. 
This attitude has a main impact on human tendency to delegating tasks to the agent so as to lessen the complexity of the task~\cite{TA}. However, both overtrust and distrust are detrimental to mission performance~\cite{HATeaming}. Overtrust can result in over-relying on the automation despite its limitations which may lead to catastrophic consequences~\cite{trustDef}. For example, in~\cite{evacuation} participants opted to follow robot instructions in emergency evacuation scenarios even when the robot provided circuitous routes to the exit. Distrust, on the other hand, may lead people to reject  the automation, hence missing its potential benefits~\cite{trustDef}. It is therefore evident that trust calibration according to the real capabilities of the swarm can be crucial to mission success and enhanced performance. 

Many factors were found to impact trust development in HRI; these factors can be related to the human (e.g. prior experience), the robot (e.g. performance), or the environment (team shared mental models). However, a quantitative meta-analysis on factors affecting trust revealed that robot performance is the foremost contributer to trust~\cite{metaTrust}.  

The persistent premise in relevant literature~\cite{trustDef}~\cite{TrustPerceptionScale}~\cite{momentTrust} is that human trust in automation is of a dynamic nature, with the dynamic component being mainly influenced by automation performance. The most widespread method for assessing human trust in automation is using surveys, e.g~\cite{foundations}~\cite{TrustPerceptionScale}. Nonetheless, the intrusive nature of surveys make them impractical for measuring real-time trust within a mission. Computational models can offer a convenient and non-intrusive way for predicting the level of trust during the mission.
Clare~\cite{ClareThesis} used system dynamics (SD) modelling to develop a model for human trust in automation in a mission with multiple UV's. In this model, trust was represented as a state that positively changes with the increase of the automation performance perceived by the human. Based on the value of trust, the rate of human interventions with automation operation is calculated such that more trust leads to a lower amount of interventions. The effect of workload on human added value was incorporated by using a workload-performance table. The model was also used in~\cite{SDTrust} to predict the performance in urban search and rescue (USAR) tasks. It was able to accurately predict the performance within 2.3\%.
A similar endeavour is found in~\cite{complianceReliance}, in which Boubin et al. used system dynamics to model human compliance and reliance behaviours as impacted by the levels of trust and stress. Other probabilistic~\cite{optimo} and linear~\cite{matrixTrust}~\cite{momentTrust} models have also been proposed for estimating real-time trust.

\section{Swarm Levels of Automation and Levels of Autonomy}
%- Possible levels of swarm involvement: 
Swarm's contributions to mission performance can be attributed to both its level of automation and autonomy~\cite{HST}. Lee et al.~\cite{trustDef} defines automation as \enquote{technology that actively selects data, transforms information, makes decisions, or controls processes}. Thus, automation refers to the capabilities of the swarm and its capacity to perform a given task. The level of automation of a swarm for a certain task can be measured using its level of human dependence, as proposed in~\cite{HST}. 
\\The level of autonomy, on the other hand, refers to the degree of freedom given to the swarm. Abbass et al.~\cite{TA} defined autonomy as \enquote{the freedom to make decisions subject to, and sometimes in spite of, environmental constraints according to the internal laws and values that govern the autonomous agent}. Mi et al.~\cite{LOA} defined four levels of autonomy for UV swarms: full autonomy, machine-oriented semi autonomy, human-oriented semi-autonomy, and  manual operation. A full autonomous swarm  is given the freedom to perform the task without any intervention from the human. However, studies showed that achieving this level of autonomy for a swarm performing real applications  is still far from reach~\cite{wearable}~\cite{meta}~\cite{LOA}. Both semi-autonomous levels of autonomy imply shared task performance. While in the machine-oriented setup the swarm performs autonomously most of the time and just informs the human of important events,  in a human-oriented setting, the swarm  often relies on human instructions for decision making. The least level of autonomy, manual operation, requires the human operator to make all the decisions and perform the actions of the swarm. 

A careful selection of the level of autonomy of a swarm is crucial for mission performance~\cite{HST}. %The effects of the selection of the level of autonomy in relation to the swarm level of automation are depicted in Fig \ref{LOA}. 
Levels of autonomy that exceeds the level of automation can lead to a performance drop due to overreliance that ignores the limitations of the swarm. On the contrary, levels of autonomy inferior to the swarm level of automation results in underutilization of the swarm, hence missing some of its benefits.  

Nevertheless, it is not only swarm characteristics that determine the appropriate level of autonomy, human factors  have their say as well. Higher levels of autonomy can be beneficial or even necessary to mitigate the high workload on the human ~\cite{meta}. Yet, such increased autonomy was criticised for its possible negative impact on SA~\cite{SAPerformance}~\cite{SAAutomation}. For instance, Gombolay et al.~\cite{teaming} found that although humans prefer highly autonomous robot teammates, their SA about team actions decrease with the increase of autonomy.  Human trust towards the swarm is also an issue to be considered, as selecting an autonomy level that exceeds the level of trust may lead the human to reject the automation.

\section{Flexible Autonomy for Effective Interaction}
Fixing the level of autonomy throughout the interaction can be problematic as it results in a rigid system that does not adapt to real-time changes~\cite{HST}. Such a setting was found to be associated with undesirable ramifications including complacency and skill degradation on the long term~\cite{HST}. Consequently, designing an interaction scheme that takes mission requirements, swarm characteristics, and human factors into consideration will likely lead to enhanced performance. 

\begin{figure*}[h]
\center
\includegraphics[scale=0.52]{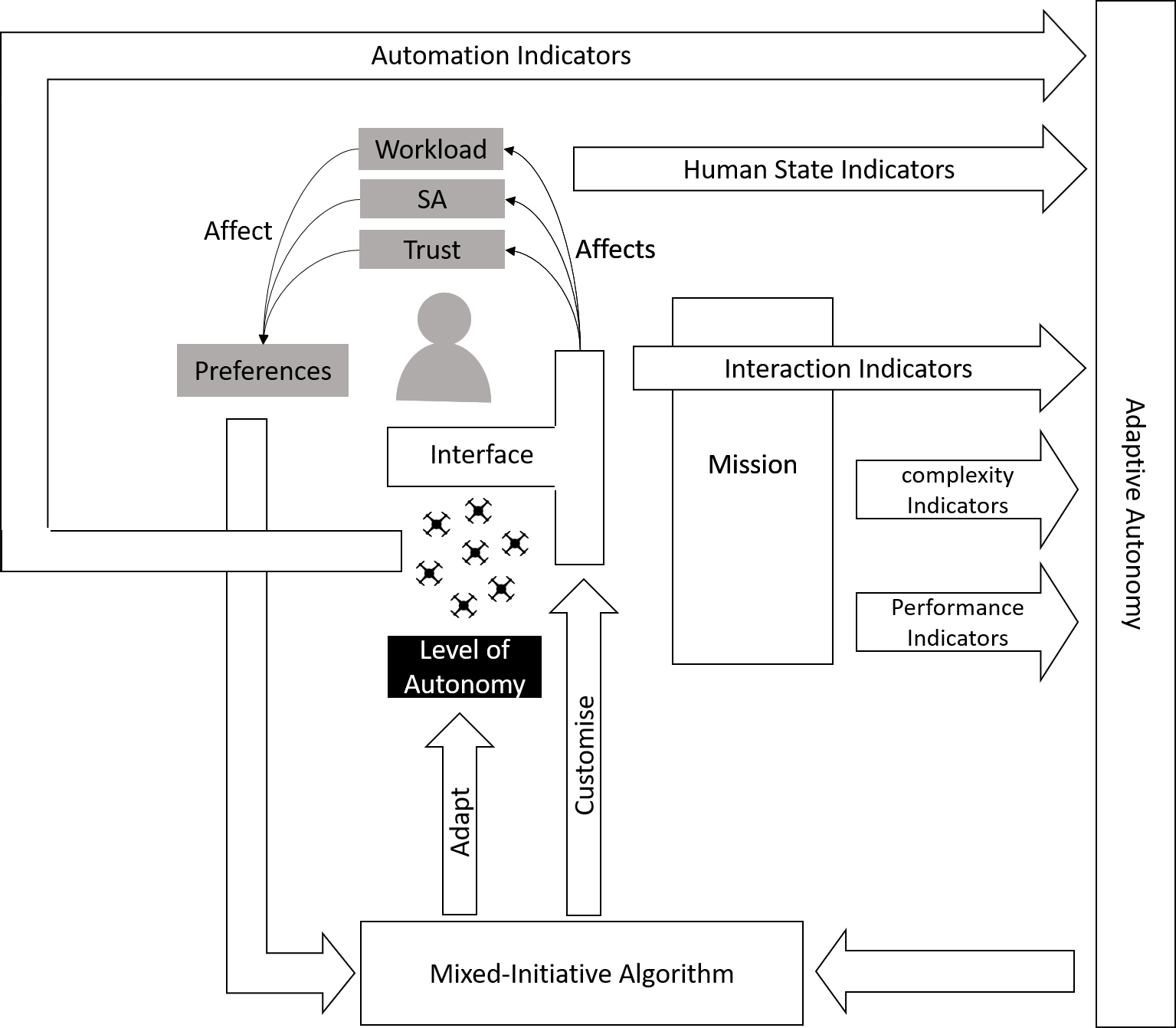}
\caption{A Mixed-initiative system for HSI.}
\label{MixedInitiative}
\end{figure*}

With flexible autonomy, smarter systems that respect the dynamics of the interaction can be achieved. Systems with flexible autonomy can invoke different levels of autonomy during the mission based on the current state.  Chen et al.~\cite{HATeaming} classified these systems into three classes based on who invokes the decision to change the autonomy:  adaptable, adaptive, and mixed-initiative systems. In adaptable systems, humans are designated to invoke the appropriate level of autonomy during the interaction. However, adaptable autonomy has been critisised for adding the workload associated with autonomy decisions to the human. On the other side, in adaptive systems these decisions are made by the automation based on its estimation of the current state so that the associated management load on the human is waived. Nevertheless, the main drawback of this arrangement is that the delegation authority is in the hands of the automation rather than the human. 
An eclectic solution that combines the advantages of the adaptable and adaptive approaches is found in mixed-initiative systems which allow for decision making that is shared between the human and the automation. These systems can integrate adaptable and adaptive components such that the adaptive component is activated under special circumstances like hard time constraints, while the adaptable component is active otherwise. An exemplar mixed-initiative system for HSI is depicted in Fig \ref{MixedInitiative}. The system is based on the work in~\cite{HST}.

% Task automation and interface customisation
Invoking a certain level of autonomy can be realised by not only applying changes to task allocation (between the human and the automation), but also by changing the interface~\cite{Feigh}.  Different interfaces were found to result in different levels of SA~\cite{SAChapter}~\cite{SAPerformance}~\cite{immersive} , workload~\cite{immersive}~\cite{cost} , trust~\cite{POMDP}~\cite{QualityTrust} , and performance %\cite{independentCamera} 
\cite{visualFeedback}~\cite{POMDP}. Elements that can be changed within the interface include display method~\cite{immersive}, interface modality and design~\cite{Feigh},  amount of information~\cite{HST}, level of information~\cite{SAChapter}, and degree of information fusion~\cite{competition} . 

The decision to change the level of autonomy of the swarm in adaptive autonomy can be triggered by relevant changes in the state of the system as evaluated by the automation. A number of research works has studied the virtues of adaptive systems triggered by the workload on the human. In~\cite{dynamicDA}, Hilburn found that workload triggered adaptive systems used in air traffic control resulted in the biggest benefit to human mental workload as it avoided very low and very high workload in low and high traffic scenarios, respectively. Abbass et al.~\cite{HusseinEEG} compared the effectiveness of workload-based adaptive systems using different techniques for evaluating workload: EEG signal, task complexity cues, or both. They found that participants perceived task complexity to be lower and their performance to be better when the workload is evaluated using EEG only than when it is evaluated using task complexity or both EEG and task complexity together. Rusnock et al.~\cite{Rusnock2} studied the effects of using different workload thresholds for invoking the adaptive autonomy on workload and SA.  They found that the proper selection of the threshold depends on the task, such that in some cases increasing the threshold can result in increasing the workload but improving SA, while in others increasing the threshold can increase the workload without benefiting SA.

Besides the workload, other indicators can also be useful in determining the proper level of autonomy. Feigh et al.~\cite{Feigh} proposed a taxonomy of triggers that can be used in adaptive systems. These triggers can be spatio-temporal-based or can be based on changes in the state of: the human, the automation, the environment, or the task. Recently, Hussein et al.~\cite{HST} identified five classes of indicators that can be used for assessing the state of the system in HSI. These classes are: performance indicators, interaction indicators, complexity indicators, swarm automation indicators, and human state indicators. Despite the potential effectiveness of these indicators in accurately evaluating the relevant state of the mission, it is not well understood how these indicators could be combined to select the proper level of autonomy. 

Unlike the question of what state indicators should be used to assess the state of the system, the question regarding the implementation details of adaptive autonomy is much less understood. Designing an effective adaptive autonomy component requires established understanding of the following issues: 
\begin{itemize}
\item How values from different indicators should be combined?
\item How can the adaptive autonomy invocation threshold be set?
\item Which tasks within the mission should be changed in its level of autonomy?
\item How can automated tasks be handed back to the human?
\item How can the interface be customised to help the human restore the SA of a previously automated task?
\item What are the effects of a certain autonomy decision on human future state?
\end{itemize}

Mixed-initiative systems including adaptive and adaptable components should be based on the understanding of some more issues as well. These issues are:
\begin{itemize}
\item When should the adaptive component be activated?
\item How to combine the adaptable and adaptive components?
\item How to inform the human of the adaptation decision to avoid undesired confusion?
\end{itemize}

The next section discusses how system modelling can contribute to this understanding.

\section{Modelling Mixed-Initiative in HSI: Challenges and Opportunities}
A number of challenges can encumber the design and implementation of a practical mixed-initiative system. We  discuss these challenges and investigate how modelling techniques can contribute to the solution.

While using a variety of indicators can lead to  more precise assessment of the state of the mission, it can lead to less effective systems than those using single indicators; e.g.~\cite{HusseinEEG}. Thus, the question of how values of different indicators can be fused to decide on whether autonomy adaptation should be triggered, becomes worthy of careful inspection. Different techniques can be applied to solve this problem. For instance, a rule-based approach can be deployed in which certain ranges of values are associated with certain autonomy levels. %However, such an approach would require high level of expertise in the details of the mission and its constituent tasks, the human factors involved, swarm behaviours, and indications of different interaction styles. 
Alternatively, machine learning approaches can be used. Both alternatives need sufficient data representing different scenarios in order to be validated and/or trained. It is theoretically possible to perform human experiments to generate the required data. However, as participants should be exposed to scenarios representing the space of possible situations, the resulting temporal and financial requirements of these experiments can make this process impractical~\cite{Rusnock1}. Additionally, with some safety critical scenarios being difficult to replicate within human experiments, the adequacy of simulated missions should be carefully assessed as they can be perceived differently by participants~\cite{evacuation}.

Understanding when adaptive autonomy should be triggered is among the first steps towards designing the adaptive component. In addition, the design of a mixed-initiative system needs to determine when the adaptive and the adaptable components should be activated. Both components can be active throughout the mission while a negotiation algorithm is used to select the final decision. Another alternative is to activate the adaptive component only during temporally critical scenarios. These design  decisions should also be based on some empirical evidence.

Once the adaptation decision is taken, changes to task autonomy, the interface, or both must be activated. This can be challenging as it depends on understanding the issues mentioned in section 5.
Unless mixed-initiative systems are designed to consider the mentioned issues, undesired ramifications can take place. For instance, a decision to fully automate a highly dynamic and mentally demanding task can relieve the high workload on the human, yet the difficulty of restoring its SA when returning it back to the human can counteract the benefits of flexible autonomy. Similarly, automating a task that requires higher capacity than the level of automation of the swarm can mitigate human's workload, however the expected performance drop may eventually lead the human to distrust the automation and hence become less willing to rely on it. Thus, an adaptation decision that is based solely on the current state of the system without taking into consideration the probable consequences can eventually lead to performance degradation.

Different modelling techniques have been successfully applied to model some aspects of HRI: workload, allocation strategy, and attention~\cite{DESWorkload}; trust, reliance, automation performance, and human performance~\cite{ClareThesis}; and workload~\cite{Rusnock1} .
These models were able to replicate data from real experiments with high accuracy, thus indicating the potential benefits of using system modelling techniques within HRI and HSI .As these models were designed to capture only some facets of the interaction, they can be used as a seed for an improved and comprehensive model that considers more aspects relevant to mixed-initiative systems, as shown in Figure~\ref{MixedInitiative}.

Models for mixed-initiative systems need to be based on a concrete understanding of the following areas: 
\begin{itemize}
\item How subjective factors like skills, cognitive abilities, experience, and personality traits can affect  human SA, workload, and trust.

\item How human factors (SA, workload, and trust) can affect and are affected by the state of the mission. 

\item Swarm levels of automation and performance for different levels of autonomy.
\item Different interaction styles humans deploy in HSI.
\item How different aspects of the interface affect human SA, workload, and trust.
\end{itemize}

Although the task of designing such a model requires considerable knowledge across different domains and is far from being trivial, its potential benefits makes the process worth it. Models for mixed-initiative autonomy would allow the designers to investigate the merits of different design decisions like how to combine human preferences and adaptive autonomy recommendations to select a certain level of autonomy, when to trigger the adaptive autonomy algorithm, and how to realise a certain level of autonomy in terms of task assignment and interface customisation. Different strategies can then be evaluated under different scenarios with respect to both the resulting mission performance and the cognitive and skill requirements on the human. 

By considering the effects of the interface on different human factors, the virtues of different interface features (e.g. levels of transparency and information fusion) can also be examined. Modelling interface issues would benefit from the literature on task complexity e.g~\cite{Campbell} and~\cite{MRT}.

Most of the proposed models for human performance capture only the objective factors (e.g. automation performance and task complexity) that affect human SA, workload, and trust. Yet, subjective factors can also lead to significant changes in human performance, as argued earlier. Although subjective factors can be harder to quantify and incorporate within a model, they can beget considerable benefits by providing insight into whether and how different people would exhibit different behaviours during a mission. Sophisticated models for SA and its relation with workload and trust can help designers examine the degree to which hiring humans with certain levels of experience and skills can be significant to mission success. 
Besides, these models make it possible to anticipate the performance gain and implications on the human state resulting from upgrading to a swarm with higher levels of automation.

\section{Conclusion}
This paper discussed open challenges that hinder the design of mixed-initiative systems in HSI. We investigated how system modelling can contribute to the solution.

Properly designed mixed-initiative HSI systems provide an effective interaction scheme that respects the dynamic nature of human factors and the capabilities of the swarm. However, the design of these systems are based on some factors that are still poorly understood, as discussed in section 5. Understanding these issues can be achieved using extensive human experiments, which can be impractical in terms of the time and monetary requirements. Fortunately, the accurate results achieved by previous models for human factors suggest that system modelling techniques can be used to establish the required understanding while minimising the need for human experiments. 

While previous systems tended to focus on some human factors while leaving others, models for mixed-initiative systems need to consider the three factors together. The adaptation decision can be based on the current levels of these factors as well as the expected effect on their future values.

As swarm characteristics, task structure, and the interface affect human factors within a mission, these aspects need to be incorporated within models for mixed-initiative systems. Requirements for these models in terms of the knowledge base and the aspects to be considered within the model have been discussed in section 6. Models satisfying these requirements would provide system designers with a handy tool that can be used to investigate a wide range of different design options and its implications on the required skills and abilities on the human. 

\section*{Acknowledgement}
This work was funded by the Australian Research Council Discovery Grant number DP140102590 and UNSW-Canberra.

%presented challenges that impede our understanding of these issues and elaborated on how modelling techniques can contribute to this understanding.
%\bibliographystyle{IEEEtranS}

%\bibliographystyle{plain}
\begingroup
\let\itshape\upshape

\mbox{}\\\endgroup
\end{document}